%% file: STM_2011_luciv.tex
\documentclass[b5paper,twoside,epsf,psfig]{stmb5_11}

\graphicspath{{../pics/}} 

\usepackage[cp1250]{inputenc}
\usepackage[T1]{fontenc}
\usepackage{euscript}
\usepackage{amsfonts,amssymb,amsmath}
\usepackage{amsthm}
\usepackage{feynmf}
\usepackage{enumerate}

\usepackage{multicol}
\usepackage{epsf}
\usepackage{hyperref}

\def\url#1{\texttt{#1}}

\newcommand{\boldnabla}{\mbox{\boldmath$\nabla$}}
\newcommand{\be}{\begin{equation}}
\newcommand{\ee}{\begin{enumerate}
  \item 
\end{enumerate}\end{equation}}
\newcommand{\ba}{\begin{eqnarray}}
\newcommand{\ea}{\end{eqnarray}}

\begin{document}

\newpage

\setcounter{equation}{0}
\setcounter{figure}{0}
\setcounter{section}{0}
\input{lucivjansky.tex}

\pagebreak

\end{document}

%% file: lucivjansky.tex


\thispagestyle{plain}
\addcontentsline{lot}{subsection}{\numberline{}\hspace*{-15mm} M.~Hnati\v{c},T.~Lu\v{c}ivjansk\'y:
 \emph{Critical Behaviour Of Directed Percolation In The Presence Of Synthetic Velocity Field}}


\STM

\title{Critical Behaviour Of Directed Percolation In The Presence Of Synthetic Velocity Field}

\authors{ M.~Hnati\v{c}$^{1,2}$, T.~Lu\v{c}ivjansk\'y$^{1,2}$}

\address{$^{1}$Institute of Experimental Physics, SAS,
Watsonova 47, 040 01 Ko\v{s}ice\\
$^{2}$Faculty of Sciences, P.~J.~\v{S}af\'arik University, Moyzesova 16, Ko\v{s}ice, Slovakia}

\bigskip

\begin{abstract}
 Using perturbative renormalization group we study the influence of
random velocity field on the critical behaviour of directed bond percolation process
 near its second-order phase transition between absorbing and active phase. We 
 consider Kraichnan model with finite correlation time for modelling advecting velocity field. Using functional
 integral representation we are able to apply field-theoretic renormalization group
 to determine possible universality classes. The model is analyzed near its 
 critical dimension by means of three-parameter expansion in $\epsilon,\delta,\eta$,
  where $\epsilon$ is the deviation from the Kolmogorov scaling, $\delta$
the deviation from the critical space dimension $d_c$ and $\eta$ is the deviation
from the parabolic dispersion law for the velocity correlator. 
Fixed points with corresponding regions of stability are evaluated to the leading order in
 the perturbation scheme. 
\end{abstract}

\section{Introduction}
Directed bond percolation (DP) problem is one of the most famous model in statistical physics exhibiting
non-equilibrium second-order phase transition \cite{Jan04}. In various formulations
it can serve for explaining hadron interactions at very high energies (Reggeon field theory) \cite{Mos78,Car80},
various models of disease spreading \cite{Jan04,Car85} or as in original formulation \cite{Broad57}
 wetting of porous material or exploring path in labyrinth. 
The upper critical dimension for this problem
was estimated to be $d=4$ in contrast to the value $d=6$ for the isotropic dynamical case \cite{Jan04,Jan85,Frey94}, which
we do not consider.
It was conjectured \cite{Jan81,Gra82} that critical regime of any one-component system
 with short-range interactions with continuos transition to absorbing state belong
  always to the DP class. Despite the general validity of this statement the percolation model
  was not experimentalluy observed until recently \cite{Tak07}. In this work transition between two topologically
different states in electrohydrodynamic convection was observed.
Diffuculties with direct experimental observation of percolation results from the fact that real material usually
 contains inhomogenities, various defects, anisotropies and so on. They can destroy or
 make observation of phase transition completely hopeless. Therefore it can be of importance to study
 possible deviations from the simple model of percolation and quantify their effects on
on percolation process. A lot of effort was put into the investigation of various  effects, e.g.
 in papers \cite{Jan99,Hin07,Jan08} long-range interactions
by the means of Levy-flight jumps were studied both in time and space variables,
 introduction of immunization was examined in \cite{Car85,Car83,Lin08}, effect of surfaces
 was studied in \cite{Jan88} etc. One can also easily imagine that spreading of disease can
 be rapidly enhanced by some external atmospheric current or by flying insects. In both
 cases because of additional drift can be modelled as random velocity field \cite{Ant08} with prescribed
 statistical properties. In this paper the influence of advective field described by rapid-change Kraichnan model was studied, which
is characterised by white-in-time nature of velocity correlator. Generalizing this approach 
it is possible to study e.g. effect of compressibility \cite{Ant10} or using stochastic Navier-Stokes
equations effects of "real" turbulent field \cite{Ant11}.
In this work we try to investigate the influence of finite correlated velocity field (for introduction see 
\cite{Ant99}) and determine how it can change the critical behaviour of percolation process. By the means of renormalization
group approach we determine possible fixed points with corresponding regions of stability. We show that 
 the model exhibit 10 possible large-scale regimes.\\
This paper is organized as follows. In section 2 we give the detailed description
of the model. In section 3 we perform dimensional analysis (power counting) of the model
and prove its multiplicative renormalizibility. In section 4 we briefly describe
main ingredients of the diagrammatic technique and present  
first order (one-loop) calculation of the renormalization constants, which are given
in the explicit form. In section 6 we analyse asymptotic behavior of the model according to its
fixed point structure. We give a comprehensive account of corresponding critical
exponents and present range of stability in the $(\epsilon,\delta,\eta)$ space.
Section 6 is devoted for conclusions and future plans.
\section{Field theoretic formulation}
The most rigorous approach to the percolation problem is based on the interpretation
of it in the means of reaction-diffusion process. According to the standard approach\cite{Jan04}
master equation for such problem can be rewritten employing Doi formalism\cite{Doi} into the
 form of time-dependent Schrodinger equation with non-hermitean hamiltonian. 
 After performing continuum limit the effective action can be derived, which
 is amenable to the usual field-theoretical methods. However it can be shown\cite{Jan04} that the phenomenological approach based on the use of 
Langevin equation with suitable chosen noise leads to the same prediction of universal quantities
 as the aforementioned master equation approach. Let us therefore briefly describe the main points 
of the latter. The stochastic non-linear differential equation for the coarse-grained density of
infected individuals(agents) $\psi(t,{\mathbf x})$ can be written in the following form
 \cite{Jan04,Tau02} 
\begin{equation}
  \partial_t \psi(t,{\mathbf x}) = D_0(\nabla^2 - \tau_0) \psi(t,{\mathbf x}) - 
  \frac{\lambda_0 D_0}{2} \psi^2(t,{\mathbf x}) + \zeta(t,{\mathbf x}),
\label{eq:lang}
\end{equation}
where $\nabla^2$ is Laplace operator, $D_0$ is a diffusion constant and $\lambda_0$ is a positive coupling constant.
 The parameter $\tau_0$ measures the deviation from the threshold value for the infection
 probability. One can assume $\tau_0\sim p_c-p$, where $p_c$ is a 
 critical probability for observing percolation (analogous to the deviation from critical temperature for
 equilibrium models).
 It is important to note, that the model has unique absorbing state with $\psi(t,{\mathbf x})=0$ 
 (no sick agents) from which it cannot escape. The Gaussian short-ranged noise $\zeta(x)$ with zero mean accounts for the density fluctuations and it has
  to respect the absorbing state condition. This can be achieved by the following choice of its
correlator\cite{Jan04,Jan99,Ant08} 
\begin{equation}
  \langle \zeta(t,{\mathbf x})\zeta(t',{\mathbf x}') \rangle = D_0 \lambda_0\psi(t,{\mathbf x}) 
  \delta(t-t')\delta({\mathbf x}-{\mathbf x}').
\label{eq:correl}
\end{equation}
In this work we want to apply renormalization group technique for the study of large scale behavior of the model (\ref{eq:lang}). It is
therefore useful to recast it into the path integral formulation. The Langevin equation (\ref{eq:lang})
has the standard form of stochastic dynamic problem\cite{Vas_book}. By introducing
Martin-Siggia-Rose\cite{Mar73} response field $\psi^\dagger(t,{\mathbf x})$ and integrating out the Gaussian noise it is possible to obtain
action functional\cite{Car80,Mar73} for the pure directed percolation problem or Reggeon field theory respectively
\begin{eqnarray}
S_{10}(\psi^\dagger,\psi) &  = & \psi^\dagger(-\partial_t + D_0 \nabla^2 - D_0\tau_0)\psi+\frac{D_0\lambda_0}{2}[(\psi^\dagger)^2\psi-\psi^\dagger\psi^2].
\label{eq:dp_act}
\end{eqnarray}
 For  convenience the required integrations over the space-time variables are not explicitly indicated
  in the action (\ref{eq:dp_act}), e.g. the second term means the following expression
\begin{equation}
  \psi^\dagger \nabla^2\psi = 
  \int dt\int d{\mathbf x}\mbox{ } \psi^\dagger(t,{\mathbf x})\nabla^2 \psi(t,{\mathbf x}).
  \label{eq:example}
\end{equation}
The model (\ref{eq:dp_act}) satisfied
 the so-called rapidity reversal symmetry (in the language of Reggeon field theory)
\begin{equation}
  \psi(t,{\mathbf x}) \rightarrow -\psi^\dagger(-t,{\mathbf x}),\quad
    \psi^\dagger(t,{\mathbf x}) \rightarrow -\psi(-t,{\mathbf x}),
\label{eq:dp_symmetry}
\end{equation}
which should be respected by the renormalization transformation. \\
In this work we would like to study the influence of advective field on the spreading of agents $\psi$. 
The agents can be considered as passive scalar quantity \cite{Ant99} that is advected by the velocity field with no back influence on the
velocity field itself with nontrivial interactions given by the cubic terms in (\ref{eq:dp_act}). 
The inclusion of the velocity field ${\mathbf v}(t,{\mathbf x})$ corresponds to the replacement
\begin{equation}
  \partial_t \rightarrow \nabla_t = \partial_t + ({\mathbf v}.\boldnabla),
  \label{eq:conv_der}
\end{equation}
in equations (\ref{eq:lang}) and (\ref{eq:dp_act}), where $\nabla_t$ is convective (Lagrangian) derivative.
The most realistic description is based on the use of stochastic Navier-Stokes equations \cite{Vas_book,Adz_book}.
  However, in this paper we shall study a simplified model in which we prescribe
statistical properties of the velocity field. Let assume that ${\mathbf v}(x)$ is a random Gaussian variable with zero mean and the
correlator given as
\begin{eqnarray}
 \langle v_i(x) v_j(x') \rangle = \int \frac{d{\bf k} d\omega}{(2\pi)^{d+1}} P_{ij}({\bf k}) D_v(\omega,{\bf k})
  \exp[-i\omega(t-t') + i{\bf k}.({\bf x}-{\bf x'})].
\label{eq:correlator_v}
\end{eqnarray}
Here $P_{ij}({\bf k}) = \delta_{ij}-k_ik_j/k^2$ is the transverse
projection operator, $k=|{\bf k}|$ is the wave number and the kernel
function $D_v$ is assumed to have the following form
\begin{eqnarray}
  D_{v}(\omega,{\bf k}) & = & \frac{g_{10} D_0^3 k^{2-2\delta-2\epsilon-\eta}}{\omega^2+u_0^2 D_0^2 (k^{2-\eta})^2}.
  \label{eq:correlator_v_fourier}
\end{eqnarray}
Here $g_{10}$ is the coupling constant (small parameter of the ordinary
perturbation theory) and the exponents $\epsilon$,$\delta$ and
$\eta$ play the role of small expansion parameters. They could be
regarded as an analog of the expansion parameter $\epsilon=4-d$ in
the usual sense of dimensional regularization. However, in this
paper $\epsilon$ should be understood as deviation of exponent of
the power law from that of the Kolmogorov scaling \cite{Frisch},
whereas $\delta$ is defined as the deviation from the space
dimension two via relation $d=4-2\delta$. The exponent $\eta$ is
related to the reciprocal of the correlation time at the wave number
$k$. The parameter $u_0$ may be used for labelling of the fixed
points and has the meaning of the ratio of velocity correlation time
and the scalar turnover time \cite{Adzhemyan2}. Although in the real
calculations $\epsilon$ is treated as a small parameter, the real
problem corresponds to the value $\epsilon=4/3$. \\
 It is interesting to note that the model for the advection field ${\mathbf v}(x)$ contains two cases of special interest:
\begin{enumerate}[(a)]
 \item in the limit $u_0 \rightarrow \infty, g_{10}' \equiv g_{10}/u_0^2 = const$ we get the 'the rapid-change model'
  $ D_v(\omega,{\bf k}) \rightarrow g_{10}' D_0 k^{-2-2\delta-2\epsilon+\eta}$,
  which is characterized by the white-in-time nature of the velocity correlator.
 \item limit $u_0 \rightarrow 0, g_{10}''\equiv g_{10}/u_0 = const$ corresponds to the case of a frozen velocity field
  $ D_{v}(\omega,{\bf k}) \rightarrow g_{10}'' D_0^2 \pi \delta(\omega) k^{2\delta-2\epsilon}$,
 when the velocity field is quenched (time-independent).
\end{enumerate}
The averaging procedure with respect to the velocity field ${\bf v}(x)$ 
may be performed with the aid of the following action functional
\begin{eqnarray}
  S_{20} =
   -\frac{1}{2}\int dt\mbox{ }dt'\mbox{ }d{\mathbf x}\mbox{ }d{\mathbf x'}
    \mbox{ } {\mathbf v}(t,{\mathbf x}) D_v^{-1}(t-t',{\mathbf x}-{\mathbf x'}) {\mathbf v}(t',{\mathbf x'}),
  \label{eq:vel_act}
\end{eqnarray}
where $D_v^{-1}$ is the inverse correlator (\ref{eq:correlator_v}) (in
the sense of the Fourier transform). The expectation value of any
relevant physical observable may be calculated using the complete
weight functional $\mathcal{W}(\psi^\dagger,\psi,{\bf v}) = {\rm
e}^{S_{10}+S_{20}}$, where $S_1$ and $S_2$ are the action functionals
(\ref{eq:dp_act}) and (\ref{eq:vel_act}).
 
The full problem is equivalent to the field-theoretic model of the four fields 
$\Phi = \{\psi^\dagger,\psi,{\mathbf v} \}$ with the total action functional given as the sum
of functionals (\ref{eq:dp_act}) and (\ref{eq:vel_act})
\begin{equation}
S_0(\Phi) =  S_{10}(\Phi) + S_{20}(\Phi).
\label{eq:total_act}   
\end{equation}
Formulation (\ref{eq:total_act}) together with (\ref{eq:vel_act}) and (\ref{eq:dp_act}) means that statistical averages of random quantities can be presented as functional averages with the weight $\exp S_0(\Phi)$ and the generating functionals of total
$G(A)$ and connected $W(A)$ Green functions are represented as functional (path) integral
\begin{equation}
  G(A) = \exp W(A) = \int \mathcal{D} \Phi \exp[S_0(\Phi)+A\Phi]
  \label{eq:gen_func}
\end{equation}
with sources $A=\{A_{\tilde{v}},A_v,A_{\psi^\dagger},A_\psi\}$ as a scalar
product $A\Phi=\sum_\Phi A_\Phi \Phi$.
All correlation and response functions can now be calculated by perturbative means in a standard fashion\cite{Vas_book,Zinn_book}.
The general consequence of causality is the vanishing of Green functions 
\begin{equation}
\langle \psi^{\dagger}(t_1,{\mathbf x}_1)\psi^{\dagger}(t_2,{\mathbf x}_2)\ldots\psi^{\dagger}(t_N,{\mathbf x}_N) \rangle,
\end{equation}
 which should be satisfied for arbitrary $N$ for any stochastic model\cite{Vas_book}. 
Let us also note that the inclusion of the velocity field does not break the symmetry (\ref{eq:dp_symmetry}).
By the direct inspection of perturbation theory it can also readily be seen that 
the real expansion parameter is rather $\lambda_0^2$ than single $\lambda_0$. This fact can also be understand 
as a consequence of the symmetry (\ref{eq:dp_symmetry}).
 Therefore we introduce new charge by the relation
\begin{equation}
  g_{20} = \lambda_0^2,
  \label{eq:new_charge}
\end{equation}
which corresponds to the charge $u_0$ used in the literature\cite{Jan04,Ant08}. 
\section{Scaling analysis and UV renormalization procedure}
The theoretical analysis of the UV divergences is based
on the power counting analysis\cite{Zinn_book}. 
In contrast to the static models quantities in dynamical models
are invariant under two independent scale transformations (with respect
to time and space variable). Therefore the canonical dimension of quantity
$Q$ is fully determined by two canonical dimensions, the frequency dimension
 $d^\omega_Q$ and momentum dimension $d^k_Q$. These dimensions are
 found are found from the usual normalization condition 
\begin{equation}
   d_\omega^\omega=-d^\omega_t = 1, d^k_k = - d^k_x = 1, d^\omega_k=d^k_\omega = 0.
  \label{eq:normal_cond}
\end{equation}
and from the requirement that the action (\ref{eq:total_act}) is dimensionless with respect
to the momentum and frequency dimensions separately. The total canonical dimension
$d_Q=d^k_Q+2d^\omega_Q$ is determined from the condition that the parabolic differential
operator of the diffusion $D_0\boldnabla^2$ and time differential operator $\partial_t$ scale uniformly under the
transformation $k\rightarrow\mu k, \omega\rightarrow \mu^2 \omega$ (corresponds to the free theory
with $\partial_t\propto D_0\boldnabla^2$).
It plays the same role as the canonical (momentum) dimension for static models.
\begin{table}[h!]   
\centering
{\begin{tabular}{cccccccccc} \hline 
\\[-1.8ex] 
$Q$ & $\psi$ & $\psi^\dagger$ & $v$ & $g_{10}$ & $\lambda_0$ & $g_{20}$ & $D_0$ & $u_0$ & $\tau_0$ 
 \\ [0.8ex] \hline
  $d_Q^k$ & $d/2$ & $d/2$ & $-1$ & $2\epsilon+\eta$ & $\delta$ & $2\delta$ & $1$ & $\eta$ & $2$ 
  \\[0.8ex] \hline
  $d_Q^\omega $& $0 $ & $0$ & $1$ & $0$ & $0$ & $0$ & $-2$ & $0$ & $0$ 
  \\[0.8ex] \hline
  $d_Q $& $d/2$ & $d/2$ & $1$ & $2\epsilon+\eta$ & $\delta$ & $2\delta$ & $0$ & $\eta$ & $2$ 
 \\[0.8ex] 
\hline \\
\end{tabular}}
\caption{Canonical dimensions of the fields and bare parameters}
\label{tab:can_dim}
\end{table}
The canonical dimensions of the model (\ref{eq:total_act}) are given in the Table \ref{tab:can_dim}. 
We see that the model is logarithmic (the canonical dimensions of the coupling
constants $g_{10},g_{20}$ and $u_{0}$ simultaneously vanish) at space dimension $d=4$ (or equivalently $\delta=0$) and 
for the choice $\epsilon=\eta0$. In what follows we will employ dimensional regularization with the minimal subtraction (MS) scheme.
According to the general theory \cite{Zinn_book} of the renormalization group the UV divergences 
in the Green functions in this scheme manifest themselves as poles in $\epsilon,\delta,\eta$ or 
possibly as their linear combination.\\
The total canonical dimension for arbitrary one-particle irreducible (1PI) Green function 
$\Gamma=\langle \Phi\ldots\Phi\rangle_{1-ir}$ is given by the relation\cite{Vas_book}
\begin{equation}
  d_\Gamma = d+2 - N_\Phi d_\Phi, 
  \label{eq:can_dim}
\end{equation} 
where $N_\Phi=\{N_v,N_{\psi^\dagger},N_\psi\}$
 are the numbers of the external fields entering into the Green function $\Gamma$
 and the summation over all types of fields is implied.
From the symmetry (\ref{eq:dp_symmetry}) it follows that the counterms corresponding to the terms
$(\psi^\dagger)^2\psi$ and $\psi^\dagger\psi^2$ can be renormalized by the same
renormalization constant. We are thus led to conclusion that all terms that should be renormalized
are already present in the action. The model (\ref{eq:total_act}) is multiplicatively renormalizable and
the renormalized action for it can be written in the general form
\begin{eqnarray}
  S_R(\Phi) & = & \psi^\dagger[-Z_1\partial_t - Z_1({\mathbf v}.\boldnabla)+Z_2 D\nabla^2 - Z_3 D\tau ]
  +\nonumber\\ & & 
  \frac{Z_4 D\lambda}{2}[(\psi^\dagger)^2\psi-\psi^\dagger\psi^2] + \frac{1}{2} {\mathbf v}D_v^{-1} {\mathbf v}.
  \label{eq:ren_act}
\end{eqnarray}

The renormalization action can be obtained by 
the multiplicative renormalization of the fields 
$\psi^\dagger\rightarrow Z_{\psi^\dagger}\psi^\dagger,\psi\rightarrow Z_{\psi}\psi$ and the parameters
\begin{eqnarray}
& &  D_0 = D Z_D, \quad g_{10} \overline{S_d} = 2g_1 \mu^{2\epsilon+\eta} Z_{g_1}, \quad \tau_0=\tau Z_\tau,
\quad u_0=u \mu^\eta Z_u,\nonumber\\
& &  \lambda_0 = \lambda \mu^\delta Z_\lambda, \quad 
  g_{20} \overline{S_d}=2g_2\mu^{2\delta}Z_{g_2},  
  \label{eq:renorm}
\end{eqnarray}
where $\overline{S_d}=S_d/(2\pi)^d$ ($S_d$ is the volume of unit sphere in $d$-dimension) is the common factor  
resulting from the momentum integration and factor $2$ was inserted for the convenience.
From the renormalized action (\ref{eq:ren_act}) and the definition (\ref{eq:renorm}) it is easy to relate renormalization
constants as follows
\begin{eqnarray}
& &  Z_1=Z_\psi Z_{\psi^\dagger} = Z_{\psi}Z_{\psi^\dagger} Z_v, \quad
     Z_2 = Z_{\psi}Z_{\psi^\dagger} Z_D,\quad
      Z_3=Z_{\psi} Z_{\psi^\dagger} Z_\tau Z_D\,,\nonumber\\
& & Z_4= Z_D Z_\lambda Z_{\psi^\dagger}^2 Z_\psi=Z_D Z_\lambda Z_{\psi^\dagger}Z_\psi^2,\quad Z_5=Z_\nu     
  \label{eq:renorm_rel}
\end{eqnarray}
Since the nonlocal term involving ${\mathbf v}$ fields in (\ref{eq:ren_act}) should not be renormalized, the relations
 $ Z_{g_1}Z_D^3=1$ and $Z_uZ_D=1$ have to be satisfied.
Inverting relations (\ref{eq:renorm_rel}) leads to the relations
\begin{eqnarray}
& & Z_v=1,\quad Z_\psi=Z_{\psi^\dagger} = Z_1^{1/2}, \quad
Z_\lambda = Z_4 Z_2^{-1} Z_1^{-1/2},\quad Z_D=Z_2 Z_1^{-1} \,,\nonumber\\
& &  Z_u=Z_1Z_2^{-1}, \quad Z_\tau = Z_3 Z_2^{-1}, Z_{g_1} = Z_1^3 Z_2^{-3} , \quad Z_{g_2}=Z_4^2 Z_1^{-1} Z_2^{-2}.
\label{eq:invert_ren}
\end{eqnarray}
\section{Calculation of the renormalization constants}
The standard perturbative approach is based on the diagrammatic expansion into the Feynman graphs\cite{Vas_book,Zinn_book}.
The inverse matrix of the free (quadratic) part of the actions (\ref{eq:dp_act}) and  (\ref{eq:vel_act})
 determines the form of the bare(unrenormalized) propagators. 
 \begin{figure}[bt]
 \centering
\includegraphics{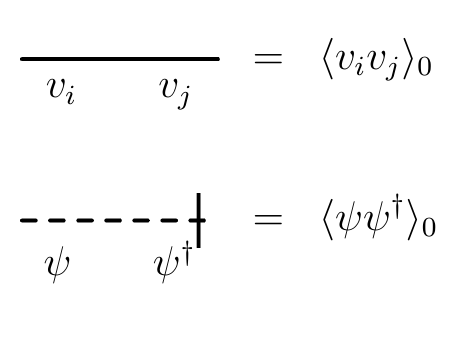}
\label{fig:propag}
\caption{The propagators of the model}
\end{figure}
In the Feynman graphs these propagators correspond to lines connecting interaction vertices. It is
 easy to see that the studied model contains three different types of propagators graphically depicted 
 in the Fig. 1. This has to be contrasted with the models\cite{Jan99,Jan08} where the inclusion
 of long-range interactions led just to the redefining $\psi^\dagger\psi$ propagator.
The propagators for the model (\ref{eq:total_act}) have the following form 
\begin{eqnarray}
 & &  \langle v v \rangle_0 = \frac{g_{10} D_0^3 k^{2-2\delta-2\epsilon-\eta}}{\omega^2+u_0^2 D_0^2 (k^{2-\eta})^2}
  P_{ij}({\mathbf k})\,,\nonumber\\
 & & \langle \psi \psi^\dagger\rangle_0=\langle \psi^\dagger\psi\rangle_0^*= \frac{1}{-i\omega_k+D_0(k^2+\tau_0)}\,,\quad
   \langle \psi \psi \rangle_0 = \langle \psi^\dagger \psi^\dagger \rangle_0 = 0
   \label{eq:prop_fre}
\end{eqnarray}
in the frequency-momentum $(\omega,{\mathbf k})$ representation.
The vertex factor\cite{Vas_book}
\begin{equation}
  V_m(x_1,x_2,\ldots,x_m;\Phi) = \frac{\delta^m V(\Phi)}{\delta\Phi(x_1)\delta\Phi(x_2)\ldots\delta\Phi(x_m)}
  \label{eq:ver_factor} 
\end{equation}
 is associated to each interaction vertex of Feynman graph.
Here, $\Phi$ could be any member from the full set of fields $\{\psi^\dagger,\psi,v \}$.
\begin{figure}[bt]
\centering
\includegraphics{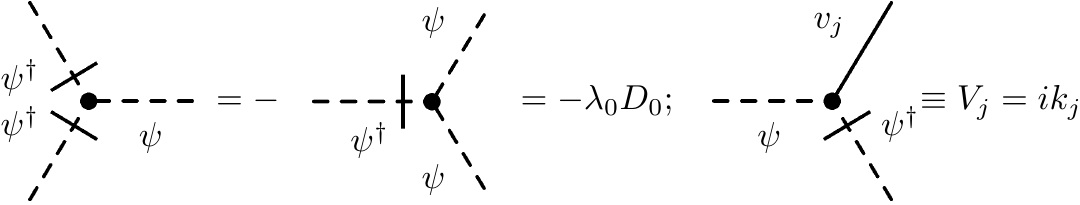}
\vspace*{8pt}
\label{fig:int_vert1}
\caption{Interaction vertices of the model}
\end{figure}
From the action (\ref{eq:dp_act}) we thus derive three possible interaction terms depicted
in the Fig. 2.
First two of them are the usual interaction vertices for direct percolation or Reggeon field theory.
The last one is responsible for the advection of spreading agent by the velocity field.
In the case of incompressible fluid it is convenient to transfer the derivation
on the response field $\psi^\dagger$
\begin{equation*}
-\int dtd{\mathbf x}\psi^\dagger \boldnabla ({\mathbf v} \psi)  =
   \int dtd{\mathbf x} \psi^\dagger \partial_i(v_i\psi)=
-\int dtd{\mathbf x}\psi^\dagger v_i\partial_i\psi=
\int dtd{\mathbf x}(\partial_i\psi^\dagger) v_i\psi,
  \label{eq:prepis_adv}
\end{equation*}
where we have used the fact that fields vanish in the infinity.
Rewriting this expression in the form $\psi^\dagger V_j v_j\psi $ we obtain immediately the vertex factor in the
momentum space 
\begin{equation}
  V_j=ik_j.
  \label{eq:int_ver_adv}
\end{equation}
From the explicit form of propagators and interaction vertices perturbation series in
Feynamn diagrams for the (connected or irreducible) Green functions could be constructed. 
From the condition of UV finitness
of 1PI functions $\Gamma_{\psi^\dagger\psi}$ and $\Gamma_{(\psi^\dagger)^2\psi}$ 
all renormalization constants $Z_1-Z_4$ can be calculated. One-loop approximation
leads in MS scheme to the following result
\begin{eqnarray}
  & & Z_1 = 1+\frac{g_2}{8\delta}, \quad Z_2 = 1+\frac{g_2}{16\delta}
  -\frac{3g_1}{8u(1+u)\epsilon},\nonumber\\
  & & Z_3 = 1 + \frac{g_2}{4\delta},\quad  Z_4 = 1+\frac{g_2}{2\delta},
  \label{eq:ren_const1}
\end{eqnarray}
from which and the relations (\ref{eq:invert_ren}) we can deduce the renormalization constants
for the fields and parameters of the model
\begin{eqnarray}
  & & Z_\tau = 1 + \frac{3g_2}{16\delta} + \frac{3g_1 }{8u(1+u)\epsilon}, \quad
  Z_\psi=Z_{\psi^\dagger} = 1 + \frac{g_2}{16\delta}, \nonumber\\
  & & Z_D = 1- \frac{3 g_1 }{8u(1+u)\epsilon} - \frac{g_2 }{16\delta}, \quad
  Z_\lambda = 1 +\frac{3 g_2 }{8\delta}+\frac{3g_1 }{8u(1+u)\epsilon}, \nonumber\\
  & & Z_{g_2} = 1+ \frac{3 g_2 }{4\delta} + \frac{3 g_1 }{4u(1+u)\epsilon}, \quad
  Z_u = 1 + \frac{g_2 }{16\delta} + \frac{3g_1}{8u(1+u)\epsilon}
 \label{eq:ren_const2}
\end{eqnarray}
\section{Fixed points}
The coefficient functions of the RG operator
\begin{align}
  & D_{{\rm RG}} = \mu\frac{\partial}{\partial\mu}\biggl|_0 = \mu\frac{\partial}{\partial \mu} +\sum_{g_i}
 \beta_{i}\frac{\partial}{\partial g_i}
  -\gamma_D D\frac{\partial}{\partial_D},
\end{align}
where the bare parameters are denoted with the subscript ``0'', are defined as
\begin{align}
  & \gamma_F=\mu\frac{\partial  \ln Z_F}{\partial_\mu}\biggl|_0 \,,\quad \beta_{i}=\mu\frac{g_i}{\mu}\biggl|_0,
\end{align}
with the charges $g_i=\{g_1,g_2,u \}$.
 From this definition and from relations (\ref{eq:renorm}) follows the explicit form for the beta functions
\begin{align}
   &  \beta_{g_1} = g_1(-2\epsilon-\eta+3\gamma_D),\hskip0.5cm \beta_u = u[-\eta+\gamma_D],\hskip0.5cm
   \beta_{g_2} = g_2(-2\delta-\gamma_{g_2})
   \label{def_beta_func}
\end{align}
and for the relevant anomalous dimensions $\gamma_D$ and $\gamma_{g_2}$
\begin{equation}
  \gamma_D = \frac{3g_1}{4u(1+u)}+\frac{g_2}{8},\quad \gamma_{g_2} =-\frac{3g_1}{2u(1+u)}-\frac{3g_2}{2}. 
  \label{eq:anomal_dim}
\end{equation}
The scaling regimes are associated with the fixed points of the corresponding RG functions. The fixed points $g^*$
 are defined as such points
$g^*=(g_1^*,u^*,g_2^*)$ for which all $\beta_g$ functions vanish
\begin{align}
  \beta_{g_1} (g_1^*,u^*,g_2^*)=\beta_u(g_1^*,u^*,g_2^*)=\beta_{g_2}(g_1^*,u^*,g_2^*)=0.
  \label{eq:fix_point}
\end{align}
The type of the fixed point is determined by the eigenvalues of the matrix $\Omega=\{\Omega_{ik}=\partial\beta_i/\partial g_k \}$,
where $\beta_i$ is the full set of $\beta$ functions (\ref{def_beta_func}) and $g_k$ is the full set of charges $\{g_1,u,g_2\}$ .
The IR asymptotic behavior is governed by the IR stable fixed points, for which all eigenvalues of $\Omega$ matrix are positive. \\
It is easy to see that the functions $\beta_{g_1}$ and $\beta_u$ satisfy relation $\beta_{g_1}/g_1 - 3\beta_u/u = 2(\eta-\epsilon)$.
This means that they cannot be equal zero simultaneously for the finite
values of the charges $g_1$ and $u$. The only exception is the instance $\epsilon=\eta$, which should be studied
separately. For general case $\epsilon\neq\eta$ we have to set either $u=0$ or $u=\infty$ and rescale $g$ in such a way, that
$\gamma_D$ remains finite \cite{Ant99}.\\
In what follows we present the results for fixed points, anomalous dimensions and eigenvalues of the $\Omega$ matrix to the first
order of perturbation theory. However, we would like to stress, that the form of $\beta$ functions (\ref{def_beta_func}) allows
to calculate the anomalous dimensions $\gamma_{D}$ and $\gamma_{\lambda}$ exactly (without any second-order correction).\\
In \cite{Adzhemyan2} independence of the renormalization constant $Z_D$ on
the exponents $\eta$ at the two-loop approximation has been conjectured. It implies that we may use the
choice $\eta=0$, which we have applied in our calculations of the renormalization constants $Z_D$
and $Z_{g_2}$. \\
Let us consider the ``rapid-change mode'' ($u\rightarrow\infty$).
It is convenient to introduce new variables $w=1/u, g_1'=g_1/u^2$ and the corresponding $\beta$ functions
obtain the form
\begin{eqnarray}
    \beta_{g_1'} = g_1'[\eta-2\epsilon+\gamma_D],\quad \beta_w = w[\eta-\gamma_D],\quad
    \beta_{g_2} = g_2(-2\delta-\gamma_{g_2}).
\end{eqnarray}
The ``rapid-change model'' corresponds to the fixed point with $w^*=0$. In this case four stable IR fixed points are realized:
\begin{eqnarray}
    \mbox{FP 1A: } & & w^*=0,\quad {g_1'}^*=0,\quad g_2^* = 0  \\
  & & \gamma_D =0,\quad \gamma_{g_2} = 0.\nonumber\\
  \mbox{stable for } & & \eta>2\epsilon,\quad \delta<0,\quad \eta>0\nonumber\\
    \mbox{FP 1B: } & & w^*=0,\quad {g_1'}^*=0,\quad g_2^* = \frac{4\delta}{3} \\
 & & \gamma_D = \frac{\delta}{6},\quad \gamma_{g_2} =- 2\delta.\nonumber\\ 
  \mbox{stable for } & & 6\eta+\delta>12\epsilon,\quad \delta>0,\mbox{ } \delta>6\eta\nonumber\\
     \mbox{FP 2A: } & & w^*=0,\quad {g_1'}^* = \frac{4(2\epsilon-\eta)}{3},\quad g_2^* = 0  \\
    & & \gamma_D=2\epsilon-\eta,\quad \gamma_{g_2}=2\eta-4\epsilon \nonumber\\
    \mbox{stable for } & & 2\epsilon>\eta>\epsilon,\quad 2\epsilon>\eta+\delta \nonumber
    \end{eqnarray}
\begin{eqnarray}  
   \mbox{FP 2B: } & & w^*=0,\quad {g_1'}^* = \frac{4(12\epsilon-6\eta-\delta)}{15},\quad 
                   g_2^* = \frac{8(\delta+\eta-2\epsilon)}{5}  \\
  & & \gamma_D=2\epsilon-\eta,\quad \gamma_{g_2}=-2\Delta \nonumber\\
 \mbox{stable for } & &      \delta>0,\quad \delta+\eta>2\epsilon,\quad 12\epsilon>6\eta+\delta \nonumber
\end{eqnarray}

For the analysis of the regime $u\rightarrow 0$ (quenched velocity field)
 we introduce the new variable ${g_1}''\equiv g_1/u$. Hence the corresponding $\beta$ functions have the form
\begin{align}
 & \beta_{{g_1}''} = {g_1}''[-2\epsilon+2\gamma_D],\quad \beta_u = u[-\eta+\gamma_D],\quad
    \beta_{g_2} = g_2(-2\delta-\gamma_{g_2})
\end{align}
Also in this case there are four possible IR stable fixed points:
\begin{eqnarray}
    \mbox{FP 3A: } & & u^*=0,\quad {g_1''}^*=0,\quad g_2^* = 0 \\
  & & \gamma_D =0,\quad \gamma_{g_2} = 0.\nonumber\\
  \mbox{stable for  }& &\epsilon<0,\quad \delta<0, \quad \eta<0 \nonumber\\
  \mbox{FP 3B: } & & u^*=0,\quad {g_1''}^*=0,\quad g_2^* = \frac{4\delta}{3} \\
   & & \gamma_D = \frac{\delta}{6},\quad \gamma_{g_2} =- 2\Delta \nonumber \\
   \mbox{stable for  } & & \epsilon<0,\quad \delta>0,\quad \eta<0 \nonumber\\
    \mbox{FP 4A: } & & w^*=0,\quad {g_1''}^* = \frac{4\epsilon}{3},\quad g_2^* = 0 \\
   & & \gamma_D=\epsilon,\quad \gamma_{g_2}=-2\epsilon \nonumber \\
   \mbox{stable for  }& &      \epsilon>0,\quad \epsilon>\delta,\quad \epsilon>\eta  \nonumber\\
   \mbox{FP 4B: } & & w^*=0,\quad {g_1''}^* = \frac{4(6\epsilon-\delta)}{15},\quad g_2^* = \frac{8(\delta-\epsilon)}{5} \\
  & & \gamma_D=\epsilon,\quad \gamma_{g_2}=-2\Delta \nonumber\\
  \mbox{stable for  }& &    0<\epsilon<\delta<6\epsilon,\quad \epsilon>\eta \nonumber
\end{eqnarray}

In the special case $\epsilon=\eta$ the functions $\beta_{g_1}$ and $\beta_u$ become proportional and this leads to the
degeneration of fixed point. Instead of just plain fixed point, we observe a whole line of fixed points in the $({g_1},u)$ plane.
\begin{eqnarray}
  \mbox{FP 5A: } & & \frac{g_1^*}{u^*(1+u^*)}=\frac{4\epsilon}{3},\quad g_2^* = 0 \\
 & & \gamma_D=\epsilon=\eta, \quad\gamma_{g_2} = -2\epsilon \nonumber\\
 \mbox{stable for  }& &  \eta=\epsilon>\delta,\quad \epsilon>0 \nonumber
\end{eqnarray}
\begin{eqnarray}  
  \mbox{FP 5B: } & & \frac{g_1^*}{u^*(1+u^*)}=\frac{4(6\epsilon-\delta)}{15},\quad g_2=\frac{8(\delta-\epsilon)}{5} \\
 & & \gamma_D=\epsilon,\quad \gamma_{g_2} = -2\delta \nonumber\\
 \mbox{stable for  }& &   \epsilon=\eta<\delta,\quad \epsilon>0.\nonumber
\end{eqnarray}  

The ``real problem'' corresponding to the Kolmogorov scaling is obtained for the value $\eta=\epsilon=4/3$, which leads to the famous ``five-thirds law'' \cite{Frisch} for the spatial velocity statistics. 
By direct observation we see that in this case critical behaviour is described by the fixed point 5A for 
logarithmic$(\delta=0,d=4)$, three dimensional$(\delta=1/2,d=3)$ and also two dimensional$(\delta=1,d=2)$ case. Fixed point 5A
is characterised by vanishing of the charge $g_2$. Therefore we conclude that in the turbulent field cubic interactions
in the vicinity of critical point (\ref{eq:dp_act}) are in fact negligible.
\section{Conclusions}
This paper is devoted to the study of directed percolation problem influenced by the external
advecting velocity field. In order to use the technique of the perturbative renormalization a field-theoretic model is constructed.
All the calculations were performed to the first order of the perturbation theory. The IR stable fixed points,
dimensions and corresponding regions of stability of fixed points are calculated.
The technically relatively simple model of velocity fluctuations used here is a convenient
starting point for more realistic high-loop calculations.
\section*{Acknowledgements}
The work was supported by VEGA grant 0173 of Slovak Academy of Sciences, and by Centre of Excellency for
Nanofluid of IEP SAS. This article was also created by implementation
of the Cooperative phenomena and phase transitions
in nanosystems with perspective utilization in nano- and
biotechnology projects No 26220120033. Funding for the operational
research and development program was provided by
the European Regional Development Fund. T.L. was sponsored by a scholarship grant by the Aktion \"Osterreich-Slowakei.